\documentclass[12pt,twoside,a4paper]{article}
\usepackage{amssymb,amsfonts,amsmath,amsthm,euscript,eufrak}
\usepackage{enumerate}
\usepackage[english]{babel}
\usepackage{dsfont}
\usepackage[latin1]{inputenc}
\usepackage{geometry}
\usepackage{graphicx}
\usepackage[bf,footnotesize]{caption}
\usepackage{sectsty}
\usepackage{mathrsfs}
\usepackage{multirow}
\usepackage{subfigure}
\usepackage{placeins}
\usepackage{lscape}
\usepackage{diagbox}
\usepackage{cite}
\usepackage{natbib} 

\sectionfont{\large}
\DeclareMathAlphabet{\mz}{OT1}{pzc}{m}{normal}
\allowdisplaybreaks

\begin{document}

\pagestyle{myheadings} % This command says to use headers.
\markboth{G.B. Cybis, M. Valk and S.R.C. Lopes}{Clustering and Classification in
DNA Sequences}

%---------------------------------------------
%  Pages size
%----------------------------------------------

\topmargin 0cm      %quanto menor o valor num\'{e}rico menor a margem superior!
\headheight 0cm        %tem rela\c{c}\~{a}o com a margem superior mas n\~{a}o sei qual \'{e} esta rela\c{c}\~{a}o...
\headsep 1.0cm         %tem rela\c{c}\~{a}o com a margem superior mas n\~{a}o sei qual \'{e} esta rela\c{c}\~{a}o...
\textheight 22.2cm     %determina a altura do texto na pagina!
\textwidth 15.0cm      %determina a largura do texto a partir da margem esquerda!
\oddsidemargin 0.6cm   %muda a largura da margem esquerda, que creio \'{e} definida pelo default!
\evensidemargin 0.3cm  %muda a largura da margem esquerda, quando imprimimos nos dois lados da folha!
\footskip 1.5cm        %determina a dist\^{a}ncia entre o fim do texto(considerando p\'{a}gina cheia) e a numera\c{c}\~{a}o da p\'{a}gina!

%----------------------------
% THEOREM Environments
%----------------------------

\theoremstyle{theorem}
\newtheorem{corollary}{\sc Corollary}[section]
\newtheorem{prop}{\sc Proposition}[section]
\newtheorem{thm}{\sc Theorem}[section]
\newtheorem{lemma}{\sc Lemma}[section]

\renewcommand\abstractname{Summary}

\theoremstyle{definition}
\newtheorem{defn}{\sc Definition}[section]
\newtheorem{remark}{\sc Remark}[section]
\newtheorem{example}{Example}[section]
\newtheorem{cond}{\sc Condition}[section]

%----------------------------
% Special commands
%----------------------------

\newcommand{\cov}{\mathrm{Cov}}
\newcommand{\corr}{\mathrm{Corr}}
\newcommand{\C}{\mathbb{C}}
\newcommand{\E}{\mathbb {E}}
\newcommand{\F}[1]{\mathcal{F}_{#1}}
\newcommand{\iten}[1]{\vspace{.2cm}\noindent{\bf #1}}
\newcommand{\K}{\mathcal{K}}
\renewcommand{\L}{\ell}
\newcommand{\LL}{\mathcal{B}}
\newcommand{\N}{\mathbb {N}}
\newcommand{\nn}[1]{\mbox{\boldmath{$#1$}}}
\newcommand{\noise}{$\{Z_t\}_{t \in \mathbb{Z}}{}$}
\renewcommand{\P}{\mathbb{P}}
\newcommand{\pe}{$\{X_t\}_{t \in \mathbb{Z}}{}$}
\renewcommand{\proof}{\noindent\textbf{Proof: }}
\renewcommand{\qed}{\begin{flushright}\vspace{-.5cm}$\Box$\end{flushright}}
\newcommand{\R}{\mathbb {R}}
\newcommand{\I}{\mathbb {I}}
\newcommand{\stg}{$\left\{X_{t}\right\}_{t=1}^{n}{}$}
\newcommand{\var}{\mathrm{Var}}
\newcommand{\Z}{\mathbb {Z}}

%----------------------------
% First page
%----------------------------
\thispagestyle{empty}

\centerline{\sc{\bf Clustering and Classification of Genetic Data Through U-Statistics\rm}}

\vspace{1.3cm}

\centerline {\sc G.B. Cybis$^a$\footnote{Corresponding author e-mail:
gabriela.cybis@ufrgs.br}, M. Valk$^a$ and S.R.C. Lopes$^a$}

\vspace{0.4cm}

\centerline{$^a$Mathematics Institute}

% \vspace{0.1cm}

\centerline{Federal University of Rio Grande do Sul}

%\vspace{0.1cm}

\centerline{Porto Alegre - RS, Brazil}

\vspace{0.4cm}

\centerline{\today}

\vspace{0.8cm}

%%%%%%%%%%%%%%%%%%%%%%%%%%%%%%%%%%%%%%%%
%   Abstract
%%%%%%%%%%%%%%%%%%%%%%%%%%%%%%%%%%%%%%%%

\begin{abstract}
%[Remove?: Many systems-biology technologies that are used to address questions in statistical genetics research, such as clustering and classification, generate high-di\-men\-sio\-nal data.]

 Genetic data are frequently categorical and have complex dependence structures that are not always well understood. For this reason, clustering and classification based on genetic data, while highly relevant, are challenging statistical problems. Here we consider a highly versatile U-statistics based approach built on dissimilarities between pairs of data points for nonparametric clustering. In this work we propose statistical tests to assess group homogeneity taking into account the multiple testing issues, and a clustering algorithm based on dissimilarities within and between groups that highly speeds up the homogeneity test. We also propose a test to verify classification significance of a sample in one of two groups.  A Monte Carlo simulation study is presented to evaluate power of the classification test, considering different group sizes and degree of separation. Size and power of the homogeneity test are also analyzed through simulations that compare it to competing methods. Finally, the methodology is applied to three different genetic datasets: global human genetic diversity,  breast tumor gene expression and Dengue virus serotypes. These applications showcase this statistical framework's ability to answer diverse biological questions while adapting to the specificities of the different datatypes. 
 \vspace{.2cm}\\
 \noindent \textbf{ Keywords.}
 Clustering, Classification, U-Statistics, Bootstrap, Genetic Data.
\end{abstract}

\vspace{0.6cm}

% \normalsize{
% %----------Classification: revisar pelo GOOGLE
% \noindent {\bf Mathematics Subject Classification (2000).} Primary
% 60G10, 62G05, 62G35, 62M10, 62M15; Secondary 62M20.}
%

%\vspace{0.4cm}

%%%%%%%%%%%%%%%%%%%%%%%%%%%%%%%%%%%%%%%%
%      Introduction
%%%%%%%%%%%%%%%%%%%%%%%%%%%%%%%%%%%%%%%%
\section{Introduction}
\renewcommand{\theequation}{\thesection.\arabic{equation}}
\setcounter{equation}{0}
\renewcommand{\thethm}{\thesection.\arabic{thm}}
\setcounter{thm}{0}
\renewcommand{\theremark}{\thesection.\arabic{remark}}
\setcounter{remark}{0}

%%% Comentarios: 
%%% Segundo comentario do Marcio, tentei colocar mais coisas sobre as aplicacoes
%%%Trocar optimization algorithm por clustering algorithm (acho que vende melhor
%%% Nao falamos nada da relaçao com ANOVA - does not depend on seccond moment assumptions... 
%%% Estamos considerando que propomos 1 ou varios testes de homogeneidade?
%%% Fiz modificaçoes apenas no abstract e intro, todas marcadas com []
%%% Ingles britânico ou americano (Tumour???)
%%% Algum conetário sobre aplicação 1 (um puco estranha)
%%% Combinar os 2 úmtimos parágrafos da intro???

The last few decades have seen a tremendous rise in the availability and diversity of genetic data, and with it, a marked increase of statistical methods tailored to answer specific biological questions. Clustering and classification are at the heart  of many of these genetic problems. In this paper we explore a model free approach for clustering and classification of genetic data based on U-statistics, that is versatile enough to be applied to a wide variety of genetic problems and adaptable enough to consider the specificities of different types of data.

%The human genome project has opened up a new page in scientific history. To this end, a variety of techniques to clustering data analysis as well as for its classification has evolved to monitor the transcript abundance for all of the organism's genes rapidly and efficiently.

Classical inference in this area generally depends on specific modeling assumptions. However, genetic data complexity present a challenge for parametric multivariate analysis techniques. In fact, details of the data generating processes are not always well understood and modeling them might involve a large number of parameters. \cite{pinheiro2005} propose an alternative method to test group homogeneity based on the Hamming distance, which gives less emphasis to the likelihood function and more to similarity measures. The test statistic is built upon comparisons of these measures between and within groups, without necessarily going through second moments. Asymptotic normality of the test statistic is obtained through properties of U-statistics. In a high-dimension low-sample-size scenario \cite{pinheiro2009} show that Hamming distance type statistics lead to a general class of first order degenerate U-statistic. Under the hypothesis of homogeneity, martingale properties are available allowing asymptotic results.
These asymptotic properties also hold without assumptions on the stochastic independence or on the homogeneity of the marginal probability laws.
%{\remove? Finally,]
Furthermore, in the work by \cite{valk_and_pinheiro_2012} these tests were adapted to the time series framework. The resulting test statistics are asymptotically Gaussian, both for the independent and identically distributed case, as well as for non-identically distributed groups of time-series under mild conditions. These conditions make it possible to deal with different correlation structures.

In this paper we explore this U-statistic clustering framework in the context of genetic data. We propose a statistical test to assess group homogeneity taking into account the issue of multiple testings. Additionally, when the Euclidean distance is considered, we present a clustering algorithm that represents significant speed-up for the homogeneity test. We also propose a test to verify the classification significance of a sample in one of two groups. We consider some simulation studies and finally we explore these results in three applications that showcase the versatility of our methods.
In the first application, we resolve small discrepancies between different tree classifications of human populations built on SNP frequency data. 
In the second one, we improve confidence in classification of a patient tumor subtype based on gene expression data through the classification test, which can lead to more reliable disease prognostics. Finally, we explore the genetic diversity of Dengue virus through sequence data, by finding genetically homogeneous clusters.

The paper is organized as follows: in Section 2 we present the basic notions of U-statistics and U-statistics based tests.
The U test for group separation, the group homogeneity test as well as the classification test are in this section. Section 3 presents a
Monte Carlo simulation study for the classification test in which we consider different sample sizes and separation degrees between the two groups to estimate
the power of the classification test. This section also contains a simulation study for comparative analysis of power and size of the homogeneity test. Section 4 presents three applications
of the methodology. In Section 5 we consider some discussions while in Section 6 we give some information of the developed software. 
In the Appendix we present the clustering algorithm and supplementary tables.

\section{U-Statistics Based Tests}\label{UstatisticsTest}
\renewcommand{\theequation}{\thesection.\arabic{equation}}
\setcounter{equation}{0}
\renewcommand{\thethm}{\thesection.\arabic{thm}}
\setcounter{thm}{0}
\renewcommand{\theremark}{\thesection.\arabic{remark}}
\setcounter{remark}{0}

 U-statistics were introduced by \cite{halmos1946} and \cite{hoeffding1948} and play an important role in estimation theory. Details
on the general theory may be found in \cite{denker1985} and \cite{lee1990}. Particularly in this work we are interested in the class of
U-statistics of order 2. For a random sample  $X_1, \cdots, X_n$ of size $n\geq 2$ sampled from a distribution $F_1$,
suppose there is a symmetric square integrable
function $g(\cdot,\cdot)$, such that $\E(g(X_1,X_2))\equiv \theta(F_1)$. Then the U-statistics with kernel $g$, defined as

\begin{equation}\label{Ustat}
 U_n=\dbinom{n}{2}^{-1}\sum_{C_{n,2}}g(X_{i_1}, X_{i_2}),
\end{equation}
\noindent
\noindent is an unbiased estimator of $\theta(F_1)$, where the above summation is over the set $C_{n,2}$ of all $(n;2)$
combinations of $2$ integers, $i_1 <i_2$, chosen from $\{1,2, \cdots,n\}$.

Consider a second random sample $Y_1, \cdots, Y_m$ of size $m\geq 2$,  drawn independently from a distribution
$F_2$ belonging to the same family of distributions as $F_1$, and let $\theta(F_1,F_2)$ be an unknown estimable parameter
in the Hoeffding's  sense \citep{hoeffding1948}. Then if there exists a function $d: \R^2 \to \R$, where $\E(d(X,Y))\equiv \theta(F_1,F_2)$,
being a distance function such as the Euclidean one, the parameter $\theta(F_1,F_2)$ will be a functional distance between
distributions $F_1$ and $F_2$. For
multivariate categorical and/or quantitative random variables where, for each $i$-th sequence, for $i \in \{1, \cdots, n\}$,
let ${\bf X}_i = (X_{i_1}, \cdots, X_{i_L})'$ be an $L$-vector and let $n$ be the sample size or the total number of sequences.
Let $\theta(F_1^{\ell}, F_2^{\ell})$ be a similar functional of the $\ell$-th marginal
distribution $F_g^{\ell}$, for $\ell = 1, \cdots, L$ and $g\in \{1,2\}$. Then assume there exists an order $2$ symmetric kernel $\phi(\cdot,\cdot)$
such that

\begin{equation}\label{eq_phi_kernel}
 \theta(F_1^{\ell},F_2^{\ell})=\int\int\phi(x_1,x_2)dF_1^{\ell}(x_1)dF_2^{\ell}(x_2).
\end{equation}
\noindent Therefore, $\theta(F_1^{\ell},F_2^{\ell})$ satisfies
\begin{equation*}
\theta(F_1^{\ell},F_2^{\ell}) \geq \frac12\,\{
\theta(F_{1}^{\ell},F_{1}^{\ell})+\theta(F_{2}^{\ell},F_{2}^{\ell})\},
\end{equation*}
\noindent for all $F_1,F_2$ and $\ell=1,\cdots,L$. If we assume that $\theta(\cdot,\cdot)$ is a convex linear function of the marginal
distributions, this implies that
\begin{equation}\label{eq_uneq}
\theta(F_1, F_2) \geq \frac12 \,\{\theta(F_1, F_1) +\theta(F_2, F_2)\},
\end{equation}
for all distributions $F_1$ and $F_2$, where equality sign holds whenever $\E(X) = \E(Y)$.

For our purpose we shall consider two groups, that is, $G=2$ (although, the theory holds
for $G\geq2$). We shall also consider two multivariate categorical and/or quantitative samples of $L$-vectors drawn from
distributions $F_1$ and $F_2$ that are  $L$-dimensional distributions defined on a common probability space.
The aim is to test the homogeneity of groups with respect to their diversity
measures. The test is based on the functional distance $\theta(\cdot, \cdot)$
as defined in \eqref{eq_uneq}, where its sample version is a generalized U-statistics.
In this multivariate setup, let $({\bf X}_{g1},\cdots ,{\bf X}_{gn_g})$ denote the vector of $n_g$ observations
in the $g$-th group of size $n_g$, for any $g \in \{1,2\}$. Therefore,

\begin{equation}\label{UnWithin}
 U_{n_g}^{(g)}=\dbinom{n_g}{2}^{-1}\sum_{1\leq i< j\leq n_g}\phi({\bf
X}_{gi},{\bf X}_{gj}),
\end{equation}
\noindent is the $g$-th generalized U-statistics, for $g \in \{1,2\}$,
with kernel $\phi({\bf x},{\bf y})$. In others words, $U_{n_g}^{(g)}$ is the estimator of the functional distance based
on distances within groups of samples drawn from the distribution $F_g$, for any $g \in \{1,2\}$.
Similarly, the generalized U-statistics
\begin{equation}\label{generalUstat}
 U_{n_1,n_2}^{(1,2)}=\frac{1}{n_1n_2}\sum_{i=1}^{n_1}\sum_{j=1}^{n_2}\phi({\bf
X}_{1i},{\bf X}_{2j})
\end{equation}
\noindent is an unbiased estimator of $\theta(F_1,F_2)$, and satisfies \eqref{eq_uneq}. \cite{pinheiro2005} consider
the following sub-group decomposition for the combined sample $U_n$

\begin{equation}\label{StatisticsUn}
 U_{n} = \sum_{g=1}^{2} \frac{n_g}{n}U_{n_g}^{(g)}+
\frac{n_1n_2}{n(n-1)}(2U_{n_1 n_2}^{(1,2)}-U_{n_1}^{(1)}-U_{n_2}^{(2)}) = W_n+B_n,
\end{equation}
\noindent where $n=n_1+n_2$ is the sample size. \cite{pinheiro2009} show that $B_n$ is in the class
of degenerate U-statistics (called quasi U-statistics) where
the asymptotic distribution is normal with convergence rates $L$ and/or $n$, even if the assumption of stochastic
independence between samples does not hold. Adapting the results in \cite{pinheiro2009} to the context of time series,
\cite{valk_and_pinheiro_2012} develop methods for classification and clustering analysis for
stationary time series.

\subsection{U test for Group Separation}\label{ClusteringTest}

We consider $G_1$ and $G_2$ two groups of samples and employ the U test for group
separation to assess whether these groups constitute statistically significant separate clusters.
Each group is assumed to be homogeneous in distribution. The null hypothesis states that both
groups are not separate, coming from the same probability distribution, while the alternative
hypothesis states that they are in fact separate groups.

The test statistics for the U test is defined as

\begin{eqnarray}\label{StatisticsBn}
 B_{n}&=&
\frac{n_1n_2}{n(n-1)}(2U_{n_1n_2}^{(1,2)}-U_{n_1}^{(1)}-U_{n_2}^{(2)}),
\end{eqnarray}
\noindent where $U_{n_1}^{(1)}$ and $U_{n_2}^{(2)}$ are the U-statistics associated to within group dissimilarities,
as defined in \eqref{UnWithin}, and $U_{n_1n_2}^{(1,2)}$ is the U-statistic associated to between group dissimilarities
as defined in \eqref{generalUstat}.

Under few regularity conditions, found in \cite{pinheiro2009}, $B_n$ is asymptotically normally distributed.
The test statistic compares weighted distances between and within groups.  Thus, from property \eqref{eq_uneq}, under
the null hypothesis, $\E(B_n)=0$, since all samples are generated from the same distribution.
Under the alternative, $\E(B_n) \geq 0$, since distances between groups are
expected to be larger than distances within groups. However, due to the fact that the
variance of $B_n$ is unknown, we employ a resampling procedure akin to permutation tests
to obtain the test statistic distribution under the null hypothesis and to assess the statistical significance
\citep{pinheiro2005}. %For all tests in this paper, the number of bootstrap (\cite{efron_and_tibshirani_1993}) samples is $1000$.

\subsection{Assessing Group Homogeneity}\label{subsec_HomTest}

The main assumption for applying the U test is homogeneity for each group.
In order to verify group homogeneity, \cite{valk_and_pinheiro_2012} employ a combinatorial procedure.
For each possible arrangement of all group elements in two subgroups,
the U test is applied. If the null hypothesis of group homogeneity is
rejected for at least one of the arrangements, then the group is considered non-homogeneous. This procedure can only
be applied if the group has at least $4$ elements, since we can only consider arrangements where each subgroup has at least two elements.

When testing in-group homogeneity for large group sizes, the number of possible assignments of all $n$ elements in 2 subgroups
(that is, $2^{n-1}-n-1$) soon becomes an important computational issue.
To reduce the computational effort of assessing overall group homogeneity we attempt to identify the subgroup configuration
that best separates the two groups. That is, if we accept the null hypothesis for the subgrouping with this configuration,
then all other arrangements will also necessarily be homogeneous. Thus, with this strategy, we need to apply the U test only once.

Note that under $H_0$ the statistic $B_n$ is asymptotically normal with zero mean. Therefore, $B_n/\sqrt{\var(B_n)}$ is asymptotically
standard normal and the group configuration that maximizes this function will also have the smallest p-value in the U test.
Thus, to test for overall group homogeneity we propose a clustering algorithm that finds the group configuration $S_1$ and $S_2$ that
minimizes the objective function

\begin{equation}\label{BnPadrTex}
 f(S_1,S_2)=\frac{-B_n}{\sqrt{\var(B_n)}},
\end{equation}
where $S_1$ and $S_2$ are the sets of observation indexes in the two groups with, respectively, $n_1$ and $n_2$ elements. Here
$n=n_1 + n_2$ is the total number of elements in $\Omega=S_1 \cup S_2$.
Then the whole group is considered heterogeneous if and only if we reject $H_0$ in the U test for this configuration.

To evaluate the objective function $f(\cdot,\cdot)$, given in (\ref{BnPadrTex}), we must estimate $\var(B_n)$.
 When $\phi(\cdot,\cdot)$ is the Euclidean distance,
 we compute the variance of $B_n$ for the independent and identically
 distributed case  (see Remark \ref{varianceBn} below). The variance of $B_n$
 under the hypothesis of group homogeneity is given by
\begin{equation}\label{eq_Var_Bn_text}
 \var(B_n)=  \frac{n_1n_2}{n^2(n - 1)^2}\left[\frac{2n^2-6n+4}{(n_1-1)(n_2-1)}\right]\sigma^4=
 C(n, n_1)\sigma^4,
\end{equation}
where $\sigma^4$, given in \eqref{eqVarpsi2}, depends only on the covariance structure of the i.i.d. vectors  $X_1,\cdots,X_n$ and $C(n, n_1)$, given in (\ref{cn}),
depends only on the overall sample size and number of elements in the first group, since $n_2=n - n_1$.

Note, however, that $\sigma^4=4\left(\mbox{vec}(\Sigma)\right)'\mbox{vec}(\Sigma)$, where $\Sigma$ is the covariance matrix
of each vector $X$ and thus has the order of $L^2$ parameters. For large values of $L$, directly estimating the variances and covariances
between the vectors components is not a feasible strategy to estimate $\var(B_n)$. We instead employ the bootstrap technique \citep{efron_and_tibshirani_1993} to estimate
$\var(B_n)$ when the size of $G_1$ is $n_1=\lfloor n/2 \rfloor$, where $\lfloor x \rfloor$ means the integer part
of $x$, and we explore the relationship between $\var(B_n)$ for different
group sizes.  If we have an estimate for the variance of $B_n$ for $n_1=i$,  $\widehat{\var_i(B_n)}$, then we can compute

\begin{equation}\label{varBnijTex}
\widehat{\var_j(B_n)}= \frac{C(n, j)}{C(n, i)}\widehat{\var_i(B_n)},
\end{equation}
for all group of size $j$, where $C(n,\cdot)$ is given by (\ref{cn}). To optimize \eqref{BnPadrTex} we employ the clustering algorithm in Appendix  A.$1$.

\begin{remark}\label{varianceBn}
In this remark we shall consider the variance of the $B_n$-statistic, defined in \eqref{StatisticsBn}.
Note that under $H_0$, $X_1, \cdots, X_n$ are i.i.d. and we assume $\mu \equiv \E (X_1)$. Additionally we note that $B_n$ is
a first-order degenerated U-statistics in the Hoeffding's sense \citep{lee1990}. Under this, we can show that
the first term in the H-decomposition is null. Thus $B_n$ can be written as a function only of the second term of this decomposition.
Let $\phi_E(X_1,X_2)=(X_1-X_2)'(X_1-X_2)$ be the Hoeffding's decomposition, where $\phi_E(\cdot,\cdot)$ is the
kernel in the expression \eqref{eq_phi_kernel} when the Euclidean distance is considered. Its conditional expectation
with respect to $X_1$ is given by
\begin{eqnarray}\label{phi1Eucl}
 \phi_1(X_1)&=&\E[(X_1-X_2)'(X_1-X_2)|X_1] = \E[X_1'X_1-2X_1'X_2+X_2'X_2|X_1] \nonumber \\
            &=&X_1'X_1-2X_1'\E[X_2|X_1]+\E[X_2'X_2|X_1] =X_1'X_1-2X_1'\mu+\E[X_2'X_2] \nonumber \\
            &=&X_1'X_1-2X_1'\mu+\mu'\mu+\mbox{tr}(\Sigma) =(X_1-\mu)'(X_1-\mu)+\mbox{tr}(\Sigma),
\end{eqnarray}
where $\Sigma$ is the covariance matrix of $X$ and $\mbox{tr}(\Sigma)$ is the trace of matrix $\Sigma$. Similarly,
the conditional expectation with respect to $X_2$ is given by
$\phi_1(X_2)=(X_2-\mu)'(X_2-\mu)+\mbox{tr}(\Sigma)$. Also $\phi_0=\E(\phi_E(X_1,X_2))$ is given by
\begin{eqnarray}\label{phi0Eucl}
 \phi_0&=&\E[(X_1-X_2)'(X_1-X_2)] = \E[X_1'X_1-2X_1'X_2+X_2'X_2] \nonumber \\
            &=&\E[X_1'X_1]-2\E[X_1'X_2]+\E[X_2'X_2]  = 2\mbox{tr}(\Sigma)+ 2\mu'\mu -2 \left(\sum_{i=1}^{d}\cov(X_{1i},X_{2i})+\mu'\mu\right)\nonumber \\
            &=&2\mbox{tr}(\Sigma),
\end{eqnarray}
and $\phi_2(X_1,X_2)=\E(\phi_E(X_1,X_2)|X_1,X_2)=(X_1-X_2)'(X_1-X_2)$. The second term of the Hoeffding's  decomposition of $\phi_E(\cdot,\cdot)$ is given by
\begin{eqnarray}\label{secHdecEucl}
\psi_2(X_1,X_2)&=&\phi_2(X_1,X_2)-\phi_1(X_1)-\phi_1(X_2)+\phi_0 = (X_1-X_2)'(X_1-X_2) \nonumber\\
&& -\{(X_1-\mu)'(X_1-\mu)+\mbox{tr}(\Sigma)\} - \{(X_2-\mu)'(X_2-\mu)+\mbox{tr}(\Sigma)\} + 2\mbox{tr}(\Sigma)\nonumber \\
            &=& (X_1-X_2)'(X_1-X_2) - (X_1-\mu)'(X_1-\mu) - (X_2-\mu)'(X_2-\mu)\nonumber \\
            &=& X_1'X_1-2X_2'X_1+X_2'X_2-X_2'X_2+2X_2'\mu -\mu'\mu-X_1'X_1+2X_1'\mu-\mu'\mu\nonumber \\
            &=&-2(X_1-\mu)'(X_2-\mu).
\end{eqnarray}

Furthermore, \cite{pinheiro2009}  say that $\psi_2(X_i,X_j)$ has the following orthogonality proprieties
\begin{equation*}
\E[\psi_2(X_i,X_j)\psi_2(X_i,X_k)] = 0 =  \E[\psi_2(X_i,X_j)\psi_2(X_k,X_l)],
\end{equation*}
\noindent for all $i,j,k, l \in \{1, 2, \cdots, n\}$. We also note that $\E[\psi_2(X_i, X_j)^2]< \infty$,
for all $i, j \in \{1, 2, \cdots, n\}$. Since the first term of Hoeffding's  decomposition is null, we can write $B_n$ as a function of
the second term $\psi_2(\cdot,\cdot)$, by the following way
\begin{eqnarray}\label{eq_Var_Bn_A}
\nonumber
 B_n &=& \frac{n_1\, n_2}{n(n-1)}\left( \frac{2}{n_1\,n_2}
\sum_{\substack{i \in S_1 \\ j \in S_2}}\psi_2(X_i,X_j) \right.\\
&& -  \left.\frac{2}{n_1(n_1 - 1)}\sum_{\substack{i,j \in S_1 \\ i<j}}\psi_2(X_i,X_j)
- \frac{2}{n_2(n_2 -1)} \sum_{\substack{i,j \in S_2 \\ i<j}}\psi_2(X_i,X_j)\right). 
\end{eqnarray}
Thus we write
{\small
\begin{eqnarray}\label{eq_Var_Bn}
\nonumber
 \var(B_n)&=& \var\left\{ \frac{n_1\, n_2}{n(n-1)}\left( \frac{2}{n_1\,n_2}
\sum_{\substack{i \in S_1 \\ j \in S_2}}\psi_2(X_i,X_j) \right. \right.\\
\nonumber
&-& \left.  \left.\frac{2}{n_1(n_1 - 1)}\sum_{\substack{i,j \in S_1 \\ i<j}}\psi_2(X_i,X_j)
- \frac{2}{n_2(n_2 -1)} \sum_{\substack{i,j \in S_2 \\ i<j}}\psi_2(X_i,X_j)\right)\right\} \\
\nonumber
&=& \frac{n_1^2\, n_2^2}{n^2(n-1)^2}\left(\frac{4}{n_1^2\,n_2^2}
\sum_{\substack{i \in S_1 \\ j \in S_2}}\var(\psi_2(X_i,X_j) )\right.\\
\nonumber
&+& \left. \frac{4}{n_1^2(n_1 - 1)^2}\sum_{\substack{i,j \in S_1 \\ i<j}}\var(\psi_2(X_i,X_j))
+\frac{4}{n_2^2(n_2 -1)^2}\sum_{\substack{i,j \in S_2 \\ i<j}}\var(\psi_2(X_i,X_j))\right) \\
\nonumber
&=& \sigma^4 \frac{n_1^2\, n_2^2}{n^2(n-1)^2} \left( \left(\frac{4}{n_1^2\, n_2^2}\right) n_1\, n_2 \,
 +  \left(\frac{4}{n_1^2(n_1 - 1)^2}\right) \frac{n_1(n_1 - 1)}{2}\right.\\
\nonumber
&+&  \left.\left( \frac{4}{n_2^2(n_2 - 1)^2}\right) \frac{n_2(n_2 - 1)}{2}\right)\\ \nonumber
&=& \sigma^4 \left(\frac{n_1^2\, n_2^2}{n^2(n-1)^2}\right)\left[\frac{4}{n_1n_2}+\frac{2}{n_1(n_1-1)}+\frac{2}{n_2(n_2-1)}\right]\\
&=& \sigma^4 \frac{n_1n_2}{n^2(n - 1)^2}\left[\frac{2n^2-6n+4}{(n_1-1)(n_2-1)}\right] = C(n,n_1)\sigma^4.
\end{eqnarray}
}
\noindent where $C(n, n_1)$ is given by
\begin{equation}
\label{cn}
C(n, n_1) = \frac{n_1n_2}{n^2(n - 1)^2}\left[\frac{2n^2-6n+4}{(n_1-1)(n_2-1)}\right]
\end{equation}
\noindent and $\sigma^4$ is given by
\begin{eqnarray}\label{eqVarpsi2}
\sigma^4&=&\var\left[\psi_2(X_1,X_2)\right] = \var\left[-2(X_1-\mu)'(X_2-\mu)\right]\nonumber\\
        &=& 4\var\left(\sum_{k=1}^{L}(X_{1k}-\mu_k)(X_{2k}-\mu_k)\right) = 4\sum_{k=1}^{L} \var \left((X_{1k}-\mu_k)(X_{2k}-\mu_k)\right)\nonumber\\
         &&   + 8\sum_{k<s}\cov\left( (X_{1k}-\mu_k)(X_{2k}-\mu_k),(X_{1s}-\mu_s)(X_{2s}-\mu_s)\right)   \nonumber\\
        &=& 4\sum_{k=1}^{L} \left(\var(X_{1k}-\mu_k)\right)^2 +8 \sum_{k<s}\cov\left( (X_{1k}-\mu_k),(X_{1s}-\mu_s) \right) ^2 = 4\mbox{vec}(\Sigma)'\mbox{vec}(\Sigma).\nonumber\\
\end{eqnarray}
\end{remark}

The procedure for assessing group homogeneity proposed by  \cite{valk_and_pinheiro_2012} involves applying the U test for
all possible group configurations. For large group sizes, when applying this strategy, we must take into account multiple testing issues.

Here, however, we propose a procedure in which only the group configuration with maximum standardized $B_n$ is tested. Thus we consider
an approximation of the distribution of $B_n$ maximum under $H_0$.

If we assume that the $B_n$'s are independent for different group configurations, then the asymptotic cumulative distribution function of the
maximum standardized $B_n$ is given by
\begin{equation}\label{eqtestmaxBn}
 F_{\mbox{max}}(x)= \P\left(\mbox{max}\left(\frac{B_n}{\sqrt{\var(B_n)}}\right)<x\right)=\Phi(x)^{\gamma},
\end{equation}
where $\Phi(\cdot)^{\gamma}$ is the standard normal cumulative distribution function at the power $\gamma$,
with $\gamma= 2^{n-1}-n -1$. If $F_{\mbox{max}}(x)>1-\alpha$, then we reject the null hypothesis of overall group homogeneity with $\alpha$ significance level.

We note that, when $\phi(\cdot,\cdot)$ is not the Euclidean distance, the variance of $B_n$ may not be constant for each group size.
Thus, the procedure described for estimating the different variances based on  (\ref{varBnijTex}) may not be applied, and we must
estimate the variance for each individual configuration with a separate bootstrap. This incurs in the same computational
effort as individual testing. Nevertheless, once the configuration with the maximum standardized $B_n$ is found, then we
can carry out the test for the maximum outlined in expression (\ref{eqtestmaxBn}), effectively correcting for multiple testings.

\subsection{Classification Test}\label{sec_ClassificationTest}

Consider the case where groups $G_1$ and $G_2$ are in fact dissimilar, as
indicated by rejection of $H_0$ in the U test.
We are interested in whether a new sample ${\bf X^*} $ would be classified in
group $G_1$ or $G_2$. \cite{valk_and_pinheiro_2012}
suggest a comparative approach based on statistics $B_1$ and $B_2$,
where $B_1$ is the statistics $B_n$ of \eqref{StatisticsBn} when the new sample is classified
in group $G_1$, and $B_2$ is defined likewise. Note that if ${\bf X^*}$ is not
well classified in $G_2$, we might expect the statistic $B_2$ to be smaller than $B_n$ computed without
including the new sample,
since this increases the distances within group $G_2$. Thus, if $B_1$ is larger
than $B_2$,
 classifying the new sample in group $G_1$ produces a better grouping in the
sense that
distances within the groups are comparatively smaller than distances between
groups.

While this procedure gives us an empirical criterion for classification, it does
not assess statistical significance.
We here propose a classification test based on the difference $D = B_1 - B_2$
to verify if the classification of ${\bf X^*}$ in group $G_1$ is statistically
significant. Let $\mu_{B_1}$ and $\mu_{B_2}$ be, respectively, the expected values of statistics
$B_1$ and $B_2$, then $\E(D)=\mu_{B_1}-\mu_{B_2}\equiv \mu_D$.
The null hypothesis states that ${\bf X^*}$ does not belong to group $G_1$,
and thus the sample arrangement that produces $B_2$ is better, or at least
equally as bad as the one that produces $B_1$.
The alternative hypothesis states that ${\bf X^*}$ is correctly classified in
group $G_1$.
The null and alternative hypotheses for this new test are given as

\begin{equation}\label{newTest}
H_0: \mu_{D} \leq 0 \ \ \mbox{versus} \ \ H_1: \mu_{D} > 0.
\end{equation}

 However, the full
distribution of $D$ is not known, and we cannot generate its distribution under the null hypothesis
through a resampling strategy. So we use the bootstrap technique \citep{efron_and_tibshirani_1993} to obtain samples from
the empirical distribution of $D$. In order to do this, for each bootstrap iteration we generate groups
$G_1^*$ and $G_2^*$ by separately resampling elements of groups $G_1$ and $G_2$ with replacement.
 We then compute the test statistic $D^*$ based on the resampled groups.
The test rejects the null hypothesis if less than $\alpha \%$ of the $D^*$ are smaller than zero, in which
case the $\alpha$ percentile of the empirical distribution of $D$ is larger than zero.

\section{Simulation Studies}
\renewcommand{\theequation}{\thesection.\arabic{equation}}
\setcounter{equation}{0}
\renewcommand{\thethm}{\thesection.\arabic{thm}}
\setcounter{thm}{0}
\renewcommand{\theremark}{\thesection.\arabic{remark}}
\setcounter{remark}{0}

In this section we present two Monte Carlo simulation studies. In the first one we analyze the size and power of the homogeneity
test proposed in Section $2.2$. In the second one we analyze the performance of the classification test proposed in Section
$2.3$.

\subsection{Size and Power of the Homogeneity Test}

We analyse the size and power of the homogeneity test proposed in Section \ref{subsec_HomTest}, to assess weather a group of samples is homogeneous. For these simulations, we consider a simple model in which the samples are sequences of length 50, generated from independent identically distributed standard multivariate normal distributions, and dissimilarities are measured by the Euclidean distance.

To study the size of the test we simulate under the null hypothesis of homogeneity, for varying group sizes $n \in \{10, 20, 40, 60\}$. We consider the homogeneity test which uses the clustering algorithm given in Appendix A.1 to find the configuration with maximum normalized U test statistic and then correct for multiple testing through the max test. We compare these results with the approach of \cite{valk_and_pinheiro_2012} of multiple U tests, and with this same approach corrected for multiple testing through the  Bonferroni correction.

\begin{table}[ht]
\caption{\normalsize{Size of the homogeneity test for different group sizes $n$.}}\label{TAB_sizeHomoTest}
\centering

\begin{tabular}{c|cccc}
\hline\hline
$n$ &  U Test   & Max Test &  Bonferroni \\
\hline\hline
  10    &   0.022   &  0.000  &    0.000         \\
  20    &   0.997   &  0.000  &    0.000         \\
  40    &   1.000   &  0.003  &    0.003         \\
  60    &   1.000   &  0.079  &    0.060         \\
\hline\hline
\end{tabular}
\end{table}

\begin{table}[ht]
\caption{\normalsize{Power of the homogeneity test for different group sizes $n_1$ and $n_2$ and different group separation.}}
\label{TAB_powerHomoTest}
\centering

\begin{tabular}{c|cccc}
\hline\hline
$n_1 \times n_2$  & U Test  & Max Test      &  Bonferroni  \\
\hline\hline

$\mu_1=0$, \,  $\mu_2=0.33$ &  \\
  \hline
   5 $\times$ 5          & 0.034    &0           &             0      \\
  10 $\times$ 10       & 0.980    &0           &             0      \\
  20 $\times$ 20       & 0.999    &0.020       &             0.018  \\ % baseado em 778 reps
  30 $\times$ 30       & 1.000    &0.465       &            0.415    \\% baseado em 522 reps
\hline
\hline
$\mu_1=0$, \, $\mu_2=0.66$ &  \\
  \hline
  5 $\times$ 5         &     0.537  &    0.001     &           0    \\
  10 $\times$ 10     &     0.995  &    0.445     &           0.386\\
   20 $\times$ 20    &     1.000  &    1.000     &           0.998\\ % baseado em 935 reps
   30 $\times$ 30    &     1.000  &    1.000     &           1.000\\
\hline
\hline
$\mu_1=0$, \, $\mu_2=1$ &  \\
 \hline
  5 $\times$ 5       &  0.994      &   0.398       &          0.312\\
  10 $\times$ 10   &  1.000      &   1.000       &          0.998\\
   20 $\times$ 20  &  1.000      &   1.000       &          1.000 \\ % baseado em 935 reps
   30 $\times$ 30   &     -       &     -         &             -  \\
  \hline\hline
\end{tabular}
\end{table}

Table \ref{TAB_sizeHomoTest} presents the size of the homogeneity test, measured as the fraction of simulations under the null hypothesis for which $H_0$ was rejected, considering the theoretical $\alpha=0.05$.  We note that the actual sizes of the tests are greatly affected by group size $n$ (even with multiple testing corrections), which is an expected consequence of the combinatorial approach that underlies our concept of homogeneity. As expected, the simple multiple U tests approach performs well only for very small groups, and if $n\geq 20$ it will almost certainly reject the null hypothesis. For the group sizes considered in this study, both the Bonferroni correction and our max test manage to control well the size.

To evaluate the power of the homogeneity test we consider a scenario in which the group is divided into two equal sized subgroups with different mean vectors. Table \ref{TAB_powerHomoTest} presents the estimated power, computed as the fraction of simulations under $H_1$ for which we reject the homogeneity hypothesis. As expected,  for all tests the power increases with group sizes and separation between groups.  Additionally, the uncorrected U test approach has higher power than the other two tests. However, since this test fails to achieve correct type 1 error probabilities in most scenarios, we would only recommend its use when the group has up to around 10 elements. We also note that the max test performs slightly better in terms of power than the Bonferroni corrected U test, in all intermediate scenarios. Thus, when the group has over 20 elements, our results favor the use of the max test.  Additional factors that favor the use of the max test in this context are the significant computational savings of 
the clustering algorithm, and the  fact that the max test arises naturally as a test for the maximum standardized U test statistic.

\subsection{Classification Test Power}

The performance of the classification test proposed in Section \ref{sec_ClassificationTest} is affected by several factors.
Critical issues are the effect of the sample size of each group and the degree of separation between groups
on the power of the test. In order to answer these issues, we perform some simulations.

The classification test can be applied to any type of data for which dissimilarity measures are available.
However, in order to perform the simulation we have to choose a specific data generating process. Due to our interest in
genetic data, and the wide use of distance methods for DNA sequences, we chose to simulate aligned DNA sequences, in a situation similar to our Dengue application (see Section \ref{sec_dengue}).
The data simulation emulates the evolution of sequences on phylogenetic trees. We first generate separate coalescent
trees for each group \citep{kingman1982} and link the trees through their roots with a branch of length $\tau$
multiplied by the root hight of the largest tree. The parameter $\tau$ is our proxy measure for the degree of
separation between groups. We then simulate the evolution of the $n_1+n_2+1$ DNA sequences along the combined
tree using the HKY base substitution model \citep{hasegawa1985}.

In order to estimate the power of the classification test we generate DNA sequences under the alternative hypothesis
that the sequence being classified ${\bf X^*}$ belongs to group $G_1$. This is done by generating group $G_1$ with
$n_1+1$ sequences and randomly assigning one as ${\bf X^*}$. We performed 1000 simulations under this scheme and
applied the separation U test to each one by using the HKY distance \citep{hasegawa1985}. In those
simulations where there is a significant separation between the $n_1$ elements of $G_1$ and the $n_2$ elements of $G_2$,
we then apply the classification test of Section \ref{sec_ClassificationTest} to assess the statistical significance of
classifying ${\bf X^*}$ in group $G_1$.   The power of the classification test is estimated as the proportion of these
simulations in which the null hypothesis is correctly rejected.

We perform simulations with varying degrees of between group separation, corresponding to values of
$\tau \in \{0.001, \, 0.5,\, 1,\, 2,\,4\}$. The considered sizes for group $G_1$ are $n_1 \in \{2,\, 5,\, 10,\, 50,\, 100\}$,
and we regard two different settings for the size of $G_2$: $n_2=n_1$ and $n_2=2n_1$. All simulated sequences are $1000$
bases long, and the overall evolutionary rate is 0.01. Tables \ref{TAB_poderGruposIguais} and \ref{TAB_poderGruposDif}
show the estimated power for all these simulations.

\begin{table}[ht]
\caption{\normalsize{Estimated power of the classification test for varying values of separation $\tau$
between groups and leading group size $n_1$, when $n_2=n_1$.}} \label{TAB_poderGruposIguais}
\centering
\begin{tabular}{r|rrrrr}
  \hline\hline
\mbox{Size} & \multicolumn{3}{r}{$\tau$}\\
  \hline
$n_1   \times n_2$  & 0.001 & 0.5 & 1 & 2 & 4 \\
  \hline
2 $  \times$\hphantom{x}  2          & 0.00 & 0.00 & 0.21 & 0.11 & 0.41 \\
5 $   \times$\hphantom{x}   5         & 0.39 & 0.64 & 0.74 & 0.90 & 0.95 \\
10 $  \times$ 10     & 0.59 & 0.78 & 0.89 & 0.95 & 0.99 \\
20 $   \times$ 20     & 0.72 & 0.86 & 0.92 & 0.97 & 0.99 \\
50 $   \times$ 50     & 0.80 & 0.90 & 0.96 & 0.99 & 0.99 \\
100 $ \times$100 & 0.86 & 0.92 & 0.95 & 0.99 & 1.00 \\
   \hline\hline
\end{tabular}
\end{table}

\begin{table}[ht]
\caption{\normalsize{Estimated power of the classification test for varying values of separation $\tau$ between
groups and leading group size $n_1$, when $n_2=2n_1$.}}\label{TAB_poderGruposDif}
\centering
\begin{tabular}{r|rrrrr}
  \hline\hline
\mbox{Size} & \multicolumn{3}{r}{$\tau$}\\
  \hline
$n_1   \times n_2$  & 0.001 & 0.5 & 1 & 2 & 4 \\
  \hline
2 $ \times$\hphantom{x}  4     & 0.01 & 0.36 & 0.62 & 0.80 & 0.92 \\
5 $ \times$\hphantom{x}10   & 0.46 & 0.70 & 0.81 & 0.92 & 0.98 \\
10 $ \times$\hphantom{x}20 & 0.64 & 0.81 & 0.88 & 0.96 & 1.00 \\
20 $\times$\hphantom{x}40 & 0.74 & 0.85 & 0.94 & 0.98 & 0.99 \\
50 $\times$100 & 0.83 & 0.91 & 0.96 & 0.99 & 1.00 \\
 \hline\hline
\end{tabular}
\end{table}

Our simulations show that increasing the number of elements in both groups leads to power increases for all scenarios.
This is a good indication of test consistency. In general, group sizes of around 5 already have reasonable power,
if the groups are well separated. By comparing Tables \ref{TAB_poderGruposIguais} and \ref{TAB_poderGruposDif},
we note that increasing the number of elements in $G_2$,
the group to which ${\bf X^*}$ does not belong, generally leads to higher power. However, for larger group sizes the impact on the
power is negligible.
Additionally, as expected, increasing the separation $\tau$ between groups also leads to power increases.
The parameter $\tau$ indicates how much more evolution occurred between the groups, in comparison with the
maximum amount of evolution inside the groups. As expected, the more separate the groups are, the easier
it is to verify that ${\bf X^*}$ in fact belongs to group $G_1$. However, even when $\tau$ is extremely low the test has
considerably high power for large group sizes. This is at least partially due to the fact that these simulations use
the U test to enforce the assumption that groups $G_1$ and $G_2$ are in fact separate, and we only consider for
power estimation purposes those cases in which the separation assumption is satisfied.

In future work, we shall consider an extensive study of the effects of dissimilarity measure choices on analyses results. It is important to understand if and how different measures may affect performances for the homogeneity and classification tests. In a different context, the work of \citep{cybis2011} compared different types of bases substitution models for DNA sequences through the likelihood ratio test. From the asymptotic theory, the authors proposed a low computational  cost estimator for the power of the likelihood ratio test.

%%%%%%%%%%%%%%%%%%%%%%%%%%%%%%%%%%%%%%%%%%%%%%%%%%%%%%%%%%%%%%%%%%%%%
%%%%%%%%%%%%%%%%
\section{Applications}\label{sec_applications}
\renewcommand{\theequation}{\thesection.\arabic{equation}}
\setcounter{equation}{0}
\renewcommand{\thethm}{\thesection.\arabic{thm}}
\setcounter{thm}{0}
\renewcommand{\theremark}{\thesection.\arabic{remark}}
\setcounter{remark}{0}

In order to showcase our methods, we now present three applications to problems of biological classification 
based on different types of genetic data.

\subsection{Global human genetic diversity}\label{sec_HGDP}
The Human Genetic Diversity Project (HGDP) is a collaboration that makes
publicly available several datasets of human genetic information.
We here consider the HGDP 2002 dataset, that contains data for 377 autossomal
microsatelite markers in 1056 individuals from 52 populations
\citep{rosenberg2002}. These data have been previously considered in different
studies to assess the evolutionary relationships among
the populations \citep*{rosenberg2002,chen_et_al_2015}. Through
alternative methodologies,
the studies produced tree representations of these relationships that agree in
broad strokes, but have
discrepancies regarding the placement of several populations. We employ our methodology to help resolve some of the points for which the previous studies
disagree.

For the dissimilarity measure in this analysis we consider the fixation index
$F_{ST}$, a commonly used differentiation measure in population genetics \citep{nei1973}. We compute pairwise
$F_{ST}$ values using the R package {\sc polysat} \citep{clark2011}.
For visualization purposes Figure 1(a)  presents a map produced from
multidimensional scaling of the dissimilarity matrix for all populations in the
dataset, in which populations are color coded according to the continent of the
origin. As expected, even this low-dimensional representation of the genetic
data shows some population separation according to geography.  In this
analysis, we highlight four sets of populations for which there were
classification discrepancies between the analyses in \cite{rosenberg2002} and \cite{chen_et_al_2015}

The first set consists of the East Asian populations represented in Figure 1(b).
The trees produced in both \cite{rosenberg2002} and \cite{chen_et_al_2015}
show separation between the populations represented in orange (group A) and
those in blue (group B), but disagree concerning the placement of the 
Japanese and Yakut populations. We first test genetic distances between groups A and B to
verify if these in fact represent two separate population clusters according to
the criteria discussed in Appendix A.$2$. We find that the separation between
these populations is non-significant (p-value of 0.154). Consequently, we do not
expect that including the Japanese and Yakut populations in either group will
lead to significant classification.  In fact, supplementary Table {\bf A1}
presents the p-values of the U test from Section \ref{ClusteringTest} for all possible
assignments of these populations, none of which are significant. Thus, we do not
assign these populations to either group in Figure 1(b), even though
there is a slight preference for including both populations in group B (U test p-value of 0.103).

Figure 1(c) presents a set of Central-South Asian populations in blue (Group C)
and a set of European and Middle Eastern populations in red (group D). Both
\cite{rosenberg2002} and \cite{chen_et_al_2015} indicate separation for
these groups, but differ regarding the placement of the Kalash and Uygur
populations. To verify if groups C and D represent statistically significant
clusters we employ the U test and obtain a p-value of 0.01,
indicating group separation. Furthermore, we apply the homogeneity test to both groups C and D,
and find that they are in fact homogeneous. We then consider all possible
placements for the Kalash and Uygur populations, and find evidence for
classifying the Uygurs in group C (p-values for these U tests
can be found in supplementary Table {\bf A2}). To assess statistical significance of this
assignment, we employ the classification test of Section \ref{sec_ClassificationTest} and obtain a p-value of 0.0471
indicating that the Uygurs should in
fact belong to group C. The classification criteria also indicates a slight
preference for classifying the Kalash population also in group C. However,  the
classification test indicates that this is a non significant assignment (p-value
of 0.364).

Two other groups that are also reliably separated in both previous studies are
the  Middle Eastern populations presented in red in Figure 1(d)  (group E) and
the European populations, presented in blue (group F). However, the studies diverge
on the placement of the Mozabites (located in North Africa but here classified as a Middle Eastern
population) whose genetic data place among the European populations in the
multidimensional scaling map. The U test for genetic separation of groups E and F
indicates that these are not significant population clusters (p-value of 0.854).
Likewise, clustering p-values for grouping the Mozabites with either population
were non-significant, although we find slight preference for classifying them in
group E (see Table {\bf A3}).

As proof of principle we choose two populations that are clearly separate: the
American populations, shown in blue in Figure 1(e) (group G), and the African
populations shown in red (group H). As expected, a testing for separation of
these groups yields a highly significant p-value of 0.001. Additionally, since
the African San population presents a troubling placement in Chen et al. (2014),
we test to see in which group it should be classified. Again, our
classification criteria easily places the African San with the other African populations
of group H (see Table {\bf A4} for all U test p-values) and the

\begin{landscape}
\begin{figure}
\begin{center}
 \includegraphics[scale=0.7]{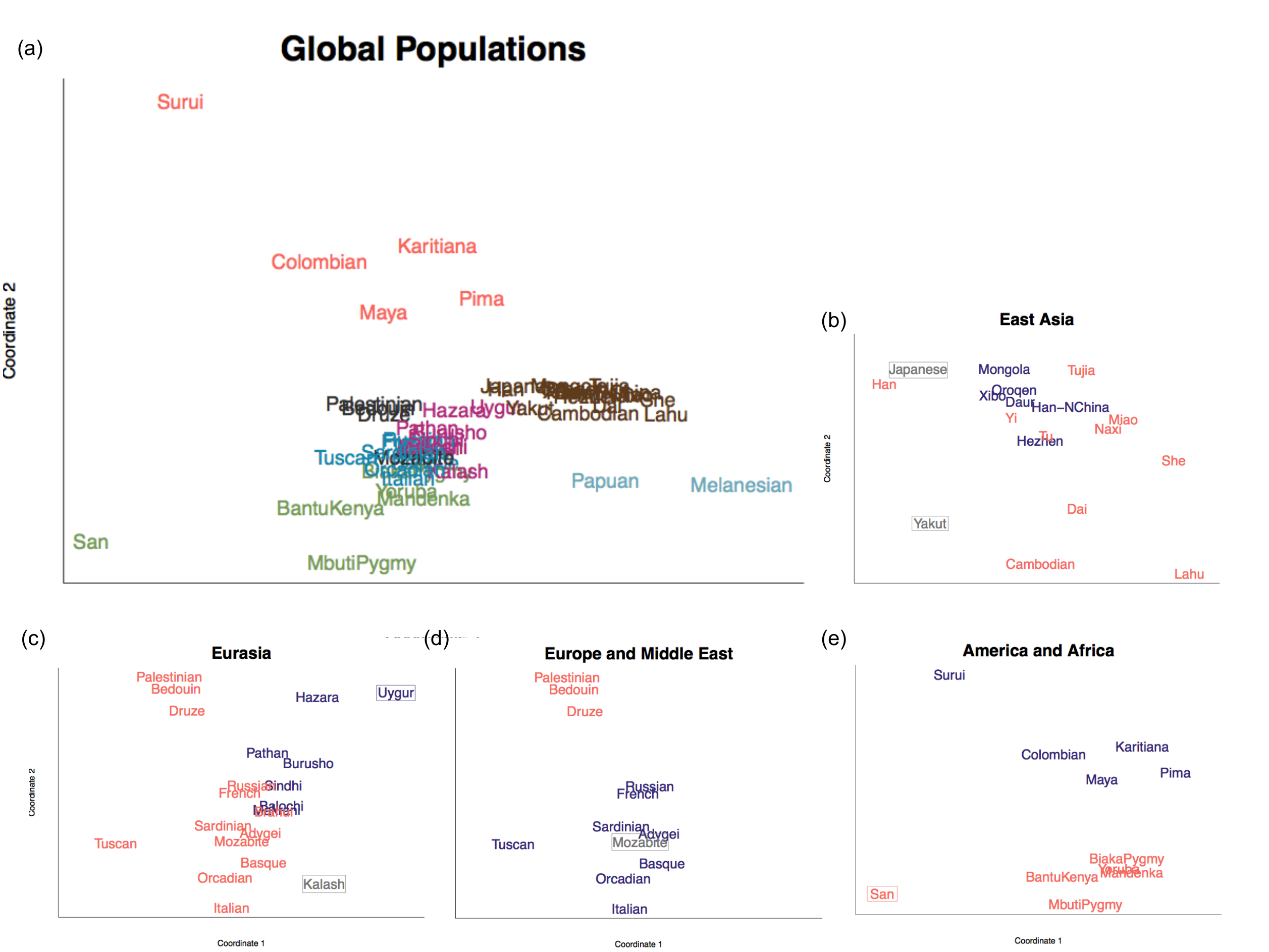}
 \caption{Multidimentional scaling maps from dissimilarity matrix. In (a) all
populations of the HGDP 2002 dataset, color coded according to region of origin:
red for Americas, green for Africa, dark blue for Europe, black for Middle East,
pink for Central-South Asia, brown for East Asia and light blue for Oceania. In
(b) East Asian populations from group A (red) and B (blue). In (c) Eurasian
populations from groups C (blue) and D (red). In (d) European and Middle Eastern
populations from groups  E (red) and F (blue). In (e) American and African
populations from groups G (blue) and H (red). Populations whose classifications
were tested are highlighted by boxes, and those in gray did not yield
significant p-values.
}
\end{center}
 \label{fig_HGPD}
\end{figure}
\end{landscape}

\noindent 
classification test indicates that this assignment is highly significant
(classification p-value $<0.0001$).

\subsection{Breast Tumor Gene Expression Clusters}\label{sec_breastCancer}

Gene expression data have been successfully used to define tumor subtypes in different types of cancer, and these
results have been associated to different clinical outcomes \citep*{ramaswamy2001,kapp2006}.
Here we analyze the Norway/Stanford dataset from \cite{sorlie2003}, that consists of gene expression data measured by DNA microarrays
for 534 genes from 122 breast tissue samples. They use machine learning techniques to classify the samples into five clinically relevant tumor
subtypes, based on gene expression profiles, and show that subtype association correlates to survival prognostics.
These genes constitute an ``intrinsic'' gene list selected by \cite{sorlie2003} as good
candidates for subtype differentiation. Their procedure consists of first selecting a few tumor samples that are archetypes
for each cluster, and then training the classification procedure on these cluster seeds.
All the 45 samples that do not belong to any cluster seed are classified  according to which subtype they fit in better.
We apply our classification test to assess whether cluster assignments are statistically significant,
potentially improving the confidence in individual prognostics.

In this application the dissimilarity measure that we use is the Euclidean distance based on expression levels for the 534 ``intrinsic'' genes.
In order to apply our methodology we must first verify if the seed samples used to define the clusters in fact constitute
distinct homogeneous groups.
The five subtypes, named Luminal A, Luminal B, Basal, ERBB2+  and Normal-like have between 10 and 27 seed elements, and were all found
to be extremely homogeneous. This was assessed using the homogeneity test with the clustering algorithm speed-up and max test correction for multiple testing.
For the clusters with larger seed groups, the speed-up of the classification algorithm is paramount
to the applicability of the homogeneity test of Section \ref{subsec_HomTest}, since in order to apply the test directly to a group of 27 elements we would need to test 67,108,836 different configurations. Given the homogeneity of all seed groups, we apply the U test to the 10 pair comparisons
between the five groups and  verify that all groups are in fact separate (with p-values $<0.002$).
Therefore, all assumptions of the classification test are satisfied.

We now wish to verify if the remaining 45 samples, that do not constitute any cluster seed, can be significantly classified
in one of the five clusters. Our classification test can only verify if the classification of a sample is statistically significant when
comparing two distinct groups. Since we want to assess significance of classification in one of the 5 groups, we adopt
the following heuristic procedure. We first compute the centroid gene expression values
for each cluster; then we verify which are the two centroids that are closer to the sample that must be classified.
This sample should be classified in the group which has the closest centroid. To assess significance of this classification,
we apply the classification test considering the two groups with closest centroids.

Of the 45 samples not assigned to any cluster seed, the classification of 18 was considered statistically significant
(with significance level $\alpha=0.05$): 13 were classified in the Luminal A cluster, 1 in the Luminal B cluster and 4
in the  Basal cluster. None of the significantly classified samples were assigned to the ERBB2+ and Normal-like clusters.

In order to evaluate the groups defined by significantly classified samples we perform a survival analysis on different
sample groups. Figure 2 presents the Kaplan-Meier analysis of relapse times, when dividing samples into 5 clusters.
First, in Figure 2(a), we consider only the samples that constitute the cluster seeds, selected for being typical examples
of each subtype. Then, in Figure 2(b), we include the seed samples and those whose classification was considered  statistically
significant. Finally, in Figure 2(c), we consider the full dataset, classifying all non-seed samples according to proximity to the cluster centroids.
We note that separation of the survival curves was improved when considering only the significantly classified samples,
in comparison to the full dataset. Additionally, the inclusion of the significantly classified samples had little effect
in subgroup survival curve separation when compared to the seed samples alone (actually, the p-value for curve separation
was slightly smaller when classified samples were included). This indicates that our method only classifies samples
that are well within the patterns of each group, obtaining group separation similar to that of only the benchmark samples.

We note, however, that while \cite{sorlie2003} employ a complex machine learning procedure to select the
optimal genes for this classification in the specific dataset, we simply consider the whole ``intrinsic" gene
list (the starting point for their analysis). For this reason, the details of our classification will differ.
Our objective here was not to provide a final classification, but to show the usefulness of our methodology in refining
the groups. Given a refined list of genes, or a set of weights for the different genes, it is straightforward to adapt our analysis.

\begin{figure}
\begin{center}
 \includegraphics[scale=0.6]{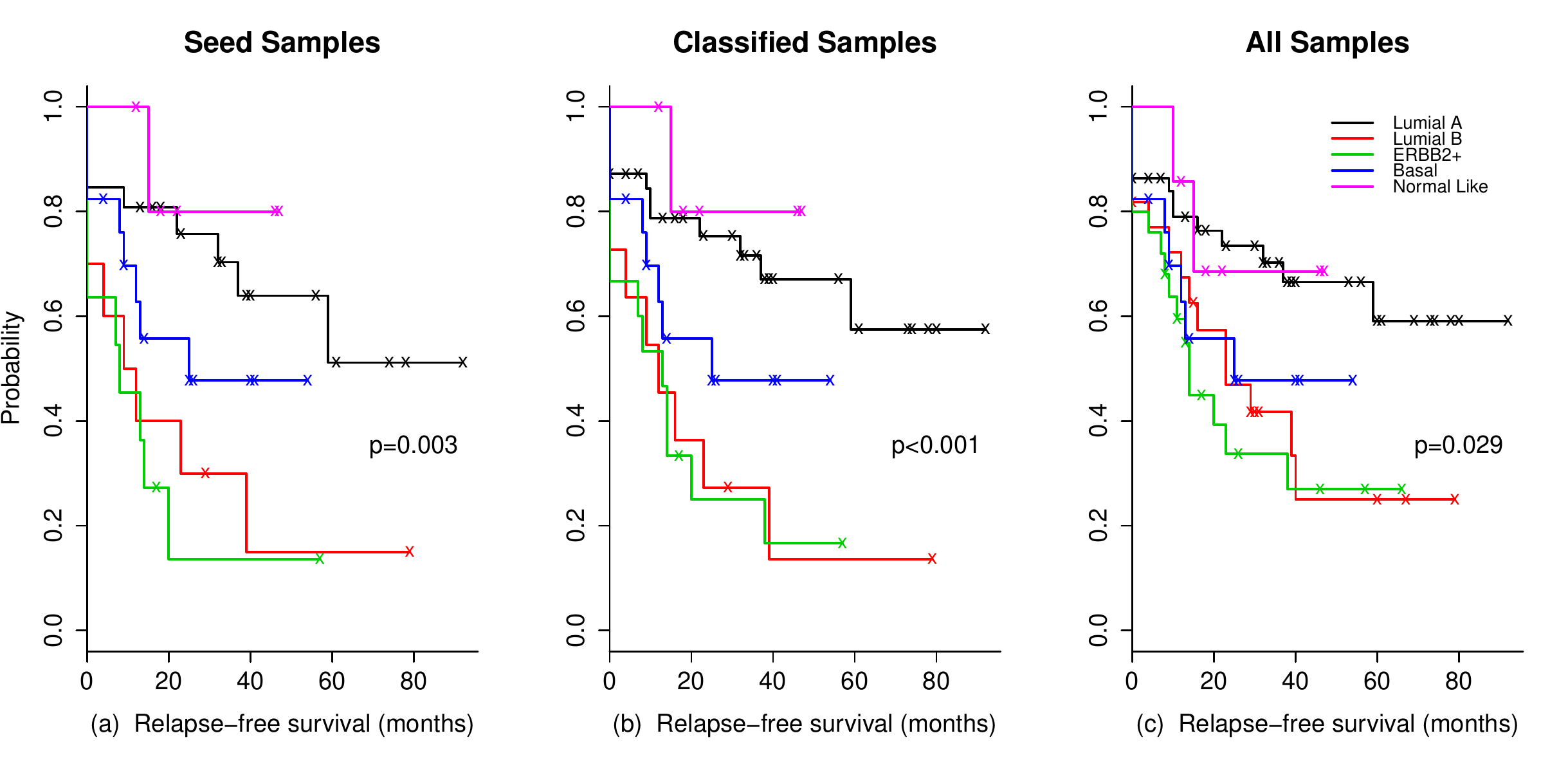}
 \caption{Kaplan-Meier analysis of relapse time. Comparing relapse times for the different clusters considering
 (a) only the samples in the cluster seeds;
 (b) cluster seed samples combined with those whose classification was considered statistically significant;
 (c) the full dataset. }
\end{center}
 \label{fig_HGPD}
\end{figure}

 %%%%%%%%%%%%%%%%%%%%%%%%%%%%%%%%%%%%%%%%%%%%%%%%%%%%%%%%%%%%%%%%%%%%%%%%%%%%%%%%%%%%%%%%%%%%%%%%%%%%%%%%%
\subsection{Dengue virus serotypes}\label{sec_dengue}

In recent years, Dengue virus has become a serious epidemiological problem in the Americas, infecting over 2 million people in 2015 alone, it is distributed in almost all countries of the continent \citep{PAHO2016}. Viral sequence data have been used, in a variety of scenarios, to study temporal, geographic and demographic aspects of rapidly evolving pathogens, such as Dengue virus \citep{pybus2012,cybis2013}.  Here we analyze the genetic variability of the virus in the Americas between the years of 2007 and 2008 by considering 144 RNA sequences from \cite{allicock2012} sampled in that period.  Our purpose in this application is not to map the whole genetic diversity of the virus (for which we would need to consider a wider range of sequences and temporal sampling), but to showcase our methods by identifying clusters of homogeneous genetic variation within the 2007 - 2008 viruses. For this analysis, we consider the HKY distance, which is built upon base substitutions, and differentiates between transition and transversion mutations \citep{hasegawa1985}.

Dengue virus has 4 phylogenetically separate serotypes DENV1 - DENV4, all represented in this sample. This is clearly reflected in the heat map of sequence distances between all samples (see Figure 3), which presents 4 clear blocks, one for each serotype. Accordingly, when we apply the homogeneity test to the whole dataset, we obtain a p-value of 0 (up to numerical precision of the Gaussian approximation), indicating a highly heterogeneous group. Additionally, pairwise U tests for group separation indicate that all serotypes in fact constitute distinct groups.

Of more interest is the structure of genetic variance within each group. Figure 4 presents Neighbour-Joining trees \citep{saitou1987} for each of the serotypes.
Applying the homogeneity test to the sequences of each individual serotype, we verify that all serotypes are composed of heterogeneous sequences ($\alpha=0.05$).

Through Figure 4-DENV1 we identify three main subgroups in Serotype 1. We applied the homogeneity test to each of the individual subgroups, and only the subgroup composed of sequences from Brazil (in brown) was considered homogeneous (p-value $= 0.8271$). Even when we remove the 2008 Nicaraguan sequence that stands out in the left group of the DENV1 tree, the group composed mainly of Mexican and Nicaraguan sequences still tests heterogeneous.

The tree for serotype 2 shows that the sequences of DENV2 for the 2007 - 2008 period are divided into two subgroups, none of which is homogeneous according to the homogeneity test. If we remove the 2007 Nicaraguan sample that stands out in the tree, the group in the upper part of the Figure 4-DENV2, composed mainly of Brazilian, Venezuelan and Colombian viruses seems to be divided into two obvious subgroups. However only the blue group of Brazilian and Puerto Rican sequences was considered homogeneous (p-value $= 0.2500$).

\begin{figure}
\begin{center}\label{fig_heatdengue}
 \includegraphics[scale=0.6]{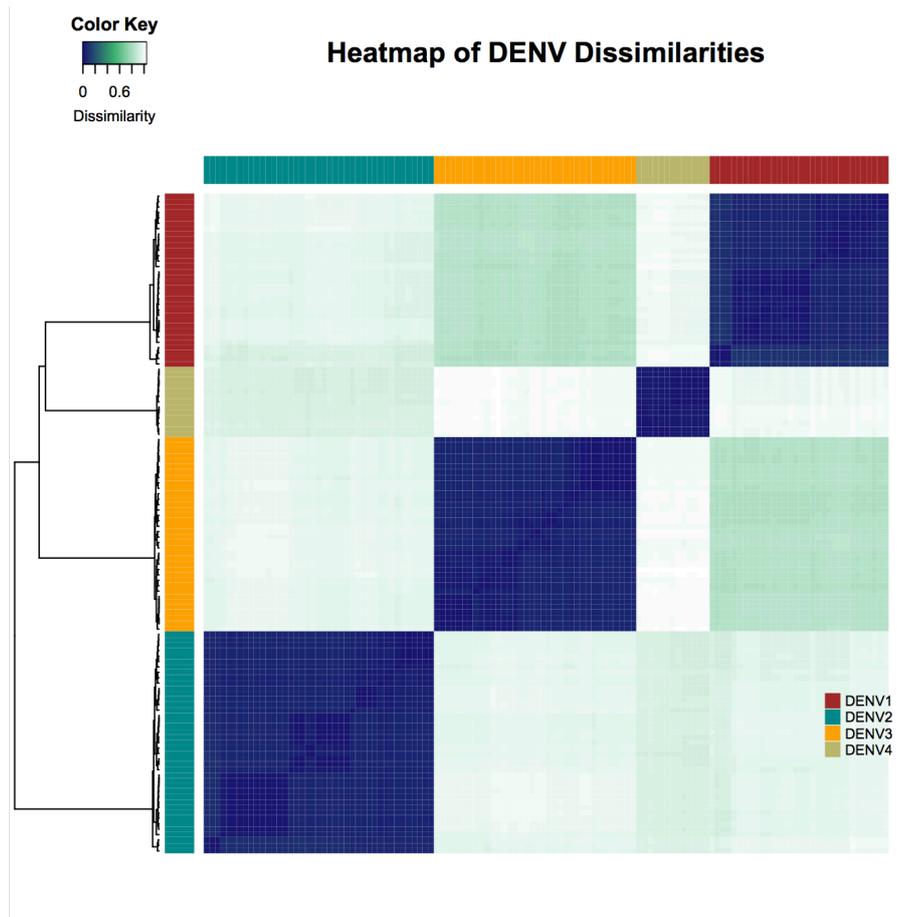}
 \caption{Heatmap of HKY distances between all DENV sequences. Side colors indicate the serotype of each sequence.}
\end{center}

\end{figure}

\begin{figure}
\begin{center}
 \includegraphics[scale=0.6]{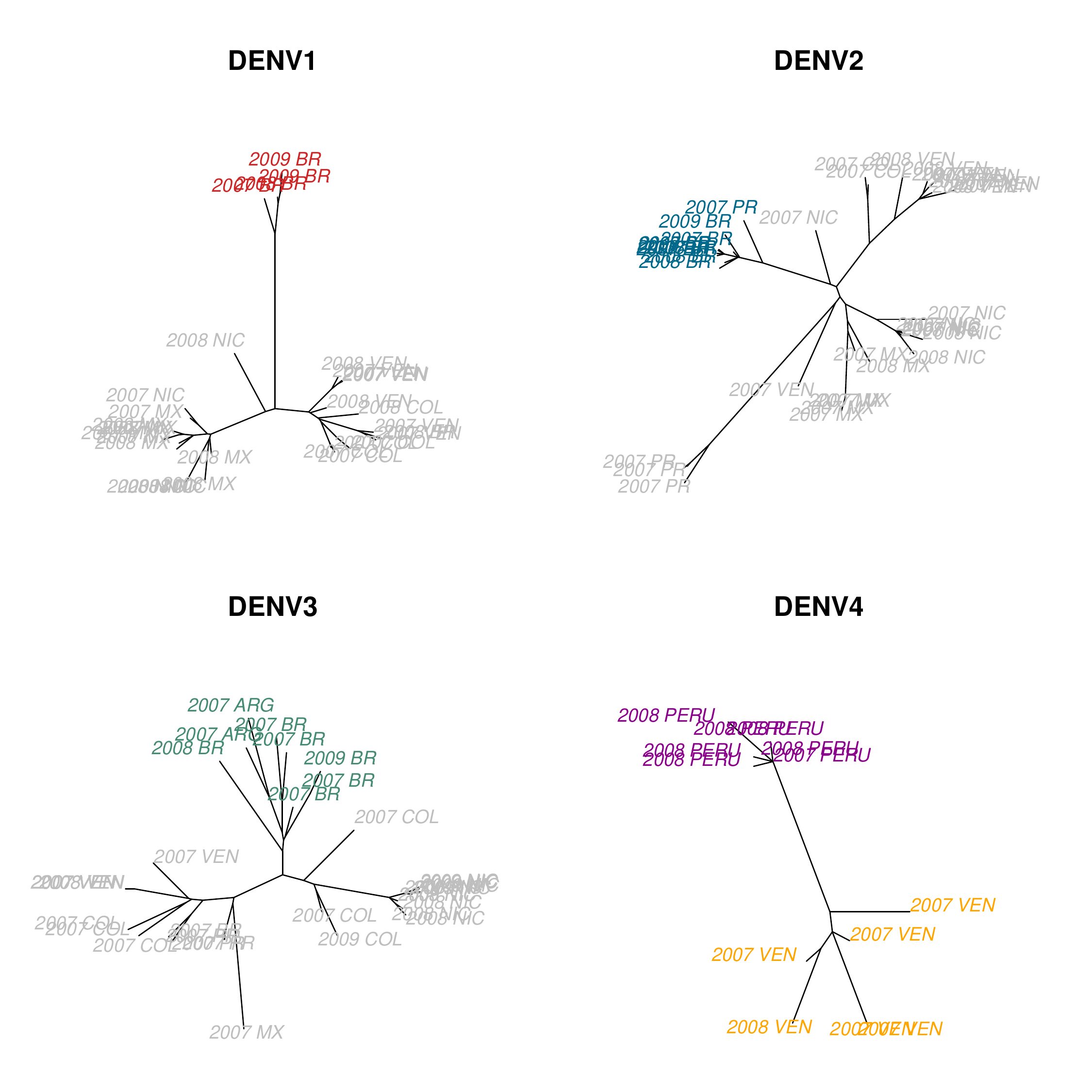}
 \caption{Neighbour-Joining tree for each Dengue virus serotype (DENV1-DENV4). Sequences are labelled according to year and country of isolation. Coloured (non-grey) labels indicate homogeneous clusters (p-value $>0.05$).}
\end{center}
 \label{fig_denvtrees}
\end{figure}

In serotype 3, we identify three major subgroups, however only the green group composed of Brazilian and Argentinian viruses was considered homogeneous (p-value $= 0.8340$). At the right side of Figure 4-DENV3, there is a group of Nicaraguan viruses that are genetically very similar, but when we applied the homogeneity test to this group, it was considered heterogeneous. This highlights the fact that homogeneity is not merely a result of small distances, but a property of the distance distributions.

Finally, the DENV4 tree divides its viruses into two subgroups. The homogeneity test confirms that both the Peruvian group (purple) and the Venezuelan group (orange) are homogeneous (see Figure 4-DENV4). Additionally the U test for group separation indicates that they are in fact distinct groups (p-value $<0.001$).

We have analyzed the genetic variation of Dengue virus between 2007  and 2008 in the Americas, and identified five homogeneous subgroups. As expected, the viruses tend to cluster according to geographical location.
We also uncovered a curious pattern in which the Brazilian sequences for serotypes 1 - 3, are all part of homogeneous groups.

%###################################################################################
\section{Discussion}

In this paper we explore U statistics based methods to solve clustering and classification problems
for genetic data in different biological settings. We propose a classification test for verifying
whether the assignment of an individual data point to one of two groups is in fact significant.
Additionally, we propose a test to assess group homogeneity, focusing on computational efficiency
and correction for multiple testing. Finally, to showcase their versatility, we apply these techniques
to three biological problems in which we address distinct clustering and classification questions using different types of genetic data.

Through the applications we exemplified how our methodology
can be used in different settings. For each dataset, we considered the appropriate dissimilarity
measure according to the peculiarities of the individual biological problems. First, in the global
human genetic diversity application of Section \ref{sec_HGDP}, we explored small discrepancies between conflicting hierarchical
classifications of human populations by assessing significance of group separation. Then, in the breast
tumor application of Section \ref{sec_breastCancer}, we sought to increase confidence in genetically based patient prognostics by assessing
significance of tumor subtype classification. Finally, in the Dengue application of Section \ref{sec_dengue}, we examined the genetic
diversity of the virus to identify clusters of genetically homogeneous strains. The versatility of these
methods is in large a consequence of their small reliance on distributional assumptions and their flexibility
in considering different dissimilarity measures.

In Section \ref{subsec_HomTest} we present the max test, a homogeneity test based on the
approach of \cite{valk_and_pinheiro_2012}. This is a test for the maximum of the standardized
U test statistic over the set of all possible subgroupings, and it arises to control for multiple
testing. Additionally, when the Euclidean distance is used, we explore the theoretical variance
of the U test statistic to build a clustering algorithm which gives significant computational
time savings. Through simulations, we established that the max test adequately controls the type I
 error as the number of elements in the group increases. Furthermore, we note that, for larger
 group sizes, the test achieves adequate power,  and we thus recommend its use for homogeneity
 testing with around 20 samples or more. For smaller group sizes, the overall type I error of
 the uncorrected multiple U test approach of \cite{valk_and_pinheiro_2012} is not largely
 affected by multiple testing, and should be preferred due to its larger power.

One use of the clustering algorithm developed for the homogeneity test, which we did not explore in this paper, is the clustering
of data into two optimal groups. Although this procedure shares conceptual similarities with $k$-means
clustering ($k=2$), it produces quite different results since it aims to simultaneously minimize
within group distances and maximize between group distances.

The Dengue application of Section \ref{sec_dengue} highlights the fact that our concept of homogeneity
is not merely a result of small distances between the samples, but a property of the distance distributions.
Thus, our method for finding genetically homogeneous groups could be applied to the study of early stages
of adaptive radiation, situation in which a group of organisms diversifies very rapidly, which may lead
specific evolutionary structures \citep{gavrilets2009}. These methods could also be employed in questions
regarding the determination of biological species  based on genetic variability.

%Discussao: It is worth noting that both \cite{rosenberg2002} and \cite{chen_et_al_2015} use geographical
%information to help inform their hierarchical clustering. We have opted to employ only the genetic
%information in our methodology, and note that

In Section \ref{sec_ClassificationTest} we explore the classification criterion of \cite{valk_and_pinheiro_2012}
for classifying a sample ${\bf X}^*$ into one of two groups to build a classification test. We employ the bootstrap to
assess significance of 2-way classification by comparing the U test statistic $B_n$, computed with ${\bf X}^*$ classified
in group $G_1$, with $B_n$ when ${\bf X}^*$ is classified in group $G_2$.
This method is tailored for a situation in which we have two reference groups, and does not naturally extend to
settings with more groups, such as the one presented in the breast cancer application of Section \ref{sec_breastCancer}.
The choice of the heuristic group centroid procedure for that 5-way classification problem reflects the centroid based
algorithm of \cite{sorlie2003}. However, we were not able to verify that this procedure satisfies some desirable
properties in 5-way classification. For instance, it is not clear that if we apply the classification test for some pair of
the 5 groups and the new sample is significantly classified in one, then it will also be significantly classified in the group
with closest centroid. This is mainly due to considerations of different group sizes. However, our heuristic procedure
presents a method for assessing statistical  significance based on a 2-way classification test that is closely related to the original problem.
In order to address these types of problems formally, an $n$-way classification test based on U statistics should be subject of future work.

Recently, \cite{clemenccon2014} also addressed the problem of clustering through U-statistics. Based on the fact that
many statistical criteria for measuring
clustering accuracy are U-statistics of order 2,  the author establishes a few bounds on clustering performance.
Through the  \emph{empirical clustering risk}, these results can be used to choose the optimal number of clusters in
an automatic model selection setting.

%For a given measure of dissimilarity between pairs of observations, the \emph{intra-cluster similarity} or the
%\emph{within cluster point scatter} is the object to be minimized. The statistical counterpart of this parameter
%(functional parameter in \cite{pinheiro2009}) is a U-statistics and is called the \emph{empirical clustering risk}.
%A lower bound is found for excess of clustering risk and an upper bound is established for the performance of
%empirical minimization of the clustering risk. These results allow to deal with the problem of automatic model
%selection, that of choose the optimal number of clusters.

Most statistical methods in genetics use simplifying assumptions on the data generating processes, and their
impacts on the analyses are not always clear. In contrast, the U statistics model free approach that we employ
here assumes only that all samples belonging to a group come from the same distribution, relying on no further
marginal distributional assumptions.

In particular, the genetic data dependency structure can be a critical modelling issue. Correlation of genetic
data within an individual genome can be a consequence of genetic linkage and functional constraints, while
correlations between samples can arise from evolutionary relatedness. In general, these processes are not
completely mapped out, and most statistical genetics methods make strong simplifying assumptions regarding
these dynamics whenever they are not the focal points of the analyses, since explicitly modeling them can
often be prohibitive. The impacts of such assumptions are not always clear. The non-parametric bootstrap
approach that we employ for the U test and the classification test implies that most of our methodology is
robust to dependency assumptions.  However, the asymptotic normality of the test statistic established in
\cite{valk_and_pinheiro_2012} depends on independence between samples and the particular choice of dissimilarity
measure. This result underlies our reasoning for the max homogeneity test. Moreover, our clustering procedure (see Section A.$1$),
that largely accelerates the homogeneity test, is built upon observations for the variance of $B_n$ derived under
the Euclidean distance as well as  between sample independence assumptions. However, at this point, the robustness
of our homogeneity test to deviations from this fairly common independence assumption has not been fully quantified. This will be the subject of future work.

%classificacao multiway - multiway tudo

%Teste de classificacao e extensao dos metodos para mais de 2 grupos. Desde o teste U, ate classificacao e homogeneidade

%\noindent ?????We will analyse the computational time issue.

%###################################################################################

\section{Software}\label{sec5}

Software in the form of R code and complete documentation is available upon
request from the corresponding author (gabriela.cybis@ufrgs.br).

%###################################################################################

\subsubsection*{Acknowledgements}
\footnotesize
S.R.C. Lopes research was partially supported by CNPq-Brazil and by INCT {\it em Matem\'atica}.

%%%%%%%%%%%%%%%%%%%%%%%%%%%%%%%%%%%%%%%%%%%%%%%%%%%%%%%%%%%%%%%%%%%%%%%%%%%%%%
%\bibliography{Classification}
%\bibliographystyle{plain}
%\bibliographystyle{ieeetr}
%\bibliographystyle{authordate1}

\normalsize{
\section*{Appendix}
\setcounter{table}{0}
\renewcommand{\thetable}{A\arabic{table}}

\subsection*{A.${\bf 1}$ Clustering  Algorithm}\label{Sec_OptmAlgor}

We used the following algorithm to find the group assignments $S_1$ and $S_2$ that minimize the objective function given in expression \eqref{BnPadrTex}.

\begin{description}
 \item[Step 1: Initialization]
\noindent
    \begin{itemize}
         \item[1.] Randomly choose starting centers for groups $G_1$ and $G_2$ from the observations $X_1, \cdots, X_n$.
         \item[2.] Assign to each observation a set of indexes, $S_1$ or $S_2$, based on the smallest Euclidean distance to its center.
     \end{itemize}
     
  \item[Step 2: Iterate] 
  \noindent
     \begin{itemize}
         \item[1.] For each observation $i \in \{1, \cdots, n\}$, assign $X_i$ to group $G_1$ if
             \[f\left(\{S_1^{-i}\cup i\},S_2^{-i}\right)<f\left(S_1^{-i},\{S_2^{-i}\cup i\}\right)\]
	      and assign $X_i$ to group $G_2$ otherwise, where $S_g^{-i}$ is the set of all indexes, except $i$,  in group $G_g$, for
$g \in \{1,2\}$.
        \item[2.] Repeat while convergence criterion is not met.
      \end{itemize}

    \item[Step 3: Convergence]
    \noindent
         \begin{itemize}
         \item[1.] Stop when $S_1$ and $S_2$ are the same in two consecutive iterations.
         \end{itemize}

\end{description}

\subsection*{A.${\bf 2}$ Supplementary Tables}

This section contains the supplementary tables for the human genetic diversity project application,
studied in Section $4.1$.

Let groups $A$ and $B$ be defined as the sets of the following populations

\begin{itemize}
\item[] $A= \{Cambodian, Dai, Han, Lahu, Miao, Naxi, Tu, Tujia, She\}$
\item[] $B=\{Daur, Han$-$NChina, Hezhen, Mongola, Oroqen, Xibo\}$.
\end{itemize}
\noindent We want to classify the $\mbox{Japanese}$ population, denoted by $J$, and $\mbox{Yakut}$ population, denoted by $Y$,
in these two Asian groups $A$ and $B$. Table ${\bf A1}$ below shows the p-values for the analysis of these two groups.

\begin{table}[ht]
\caption{\normalsize{p-Values for the analysis of groups A and B.}}\label{TAB_pValuesAB}
\centering
\begin{tabular}{ccc}
\hline\hline
  Group A & Group B & p-value \\
\hline\hline
       A    &     B        & 0.154   \\
       A+ J &     B        & 0.167   \\
       A    &    B+J       & 0.120   \\
       A    &    B+Y       & 0.122   \\
       A+Y  &     B        & 0.211   \\
       A+J+Y&     B        & 0.253   \\
       A    &     B+J+Y    & 0.103   \\
        \hline\hline
\end{tabular}

J = Japanese population; Y = Yakut population.
\end{table}

Let groups $C$ and $D$ be defined as the sets of the following populations

\begin{itemize}
\item[] $C= \{Balochi, Burusho, Hazara, Makrani, Pathan, Sindhi\}$
\item[] $D=\{Adygei, Basque, Bedouin, Brahui, Druze, French, Italian, Mozabite,$
\newline
\hphantom{xxxxx}
$Orcadian, Palestinian, Russian, Sardinian, Tuscan\}$.
\end{itemize}
\noindent We want to classify the $\mbox{Kalash}$ population, denoted by $K$, and $\mbox{Uygur}$ population, denoted by $U$,
in these two groups $C$ and $D$. Table ${\bf A2}$ below shows the p-values for the analysis of these two groups.

\begin{table}[ht]
\caption{\normalsize{p-Values for the analysis of groups C and D.}}\label{TAB_pValuesCD}
\centering
\begin{tabular}{ccc}
\hline\hline
  Group C  & Group D & p-value \\
\hline\hline
       C           &          D             & 0.010  \\
     C + K         &         D              & 0.012  \\
       C           &      D + K             & 0.023  \\
    C + U          &      D + K             & 0.007  \\
     C + K         &      D + U             & 0.018  \\
    C + U          &         D              & 0.002  \\
       C           &       D + U            & 0.036  \\
       C           &       D + K+U          & 0.041  \\
 C + K+U           &          D             & 0.007  \\
   \hline\hline
\end{tabular}

K = Kalash population; U = Uygur population.
\end{table}

Let groups $E$ and $F$ be defined as the sets of the following populations

\begin{itemize}
\item[] $E= \{Bedouin, Druze, Palestinian\}$
\item[] $F=\{Adygei, Basque, French, Italian, Orcadian, Russian, Sardinian, Tuscan\}$.
\end{itemize}
\noindent We want to classify the $\mbox{Mozabite}$ population, denoted by $M$, within the European and Middle Eastern populations.
Table ${\bf A3}$ below shows the p-values for the analysis of these two groups.

\begin{table}[ht]
\caption{\normalsize{p-Values for the analysis of groups E and F.}}\label{TAB_pValuesEF}
\centering
\begin{tabular}{ccc}
\hline\hline
  Group E &  Group  F  & p-value \\
\hline\hline
        E      &         F            & 0.854   \\
        E + M  &        F             & 0.728   \\
        E      &         F + M        & 0.872   \\
         \hline\hline
\end{tabular}

M = Mozabite population.
\end{table}

Let groups $G$ and $H$ be defined as the sets of the following populations

\begin{itemize}
\item[] $G= \{Colombian, Karitiana, Maya, Pima, Surui\}$
\item[] $H=\{BantuKenya, BiakaPygmy, Mandenka, MbutiPygmy, Yoruba\}$.
\end{itemize}
\noindent We want to classify the $\mbox{San}$ population, denoted by $S$, within the American or African populations.
Table ${\bf A4}$ below shows the p-values for the analysis of these two groups.

\begin{table}[ht]
\caption{\normalsize{p-Values for the analysis of groups G and H.}}\label{TAB_pValuesGH}
\centering
\begin{tabular}{ccc}
\hline\hline
Group G  & Group H & p-value \\
\hline\hline
    G         &        H       & 0.001     \\
  G +S      &        H       & 0.012     \\
      G       &        H +S  & 0.002     \\
      \hline\hline
\end{tabular}

S = San Population 
\end{table}

}

\end{document}